# An Overview of 5G System Accessibility Differentiation and Control

Jingya Li, Demia Della Penda, Henrik Sahlin, Paul Schliwa-Bertling, Mats Folke, Magnus Stattin

Ericsson

Emails: {jingya.li, demia.della.penda, henrik.sahlin, paul.schliwa-bertling, mats.folke, magnus.stattin}@ericsson.com

*Abstract*—5G system is characterized by its capability to support a wide range of use cases and services. Supporting accessibility differentiation becomes therefore essential to preserve a stable network condition during high traffic load, while ensuring connection and service quality to prioritized devices and services. In this article, we describe some use cases and requirements that impact the 3GPP design of the 5G accessibility differentiation framework. We then provide an overview of the supported mechanisms for accessibility differentiation and control in 5G Stand Alone (SA) system, including cell barring and reservation, unified access control, paging control, random access control and admission control. We discuss how these functionalities can be used to maintain the service quality and selectively limit the incoming traffic to the network at high load situations, leveraging different connection-type indicators and connection-priority identifiers.

## I. INTRODUCTION

The 3rd Generation Partnership Project (3GPP) has dedicated massive efforts to develop and evolve the fifth generation (5G) system to be capable of simultaneously supporting a variety of use cases, such as enhanced mobile broadband, massive machine type communications, and ultra-reliable low-latency communications [1]. The first 3GPP 5G standard was finalized in Rel-15, making 5G commercial readiness. Built on the first release, enhancements were made in Rel-16 to improve the 5G capabilities to support new deployment scenarios and new verticals, like fixed wireless access (FWA), industrial internet of things (IoT) and vehicle to everything (V2X) communications. The ongoing Rel-17 is working on further expanding the availability and the applicability of 5G system for automotive, public safety, augmented/virtual reality, and manufacturing use cases.

The flexible design of the 5G wireless access technology, i.e., new radio (NR), provides a solid foundation to meet the requirements of this variety of use cases, leveraging also the introduction of new concepts like network slicing [2], [3]. Because of the diverse service requirements and the limitation of network resources, to realize the full potential of NR, it is crucial to design mechanisms that control the amount of traffic accessing the network and thus consuming resources. This is to prevent congestion and network malfunction or crash, which can lead to connection drops and network unavailability for prioritized, and even vitally important, requests like emergency calls or critical communication between first responders.

Network accessibility control becomes important for preserving a stable working condition of the network, for maximizing the number of simultaneous transmissions with satisfactory performance. It is also an essential 5G technology component to provide the desired accessibility prioritization for different types of devices, users and services, when supporting multiple verticals in a shared 5G Stand Alone (SA) network, as illustrated in Figure 1.

This article provides first an overview of the motivations for accessibility differentiation in 5G system. Then, it describes the mechanisms introduced by 3GPP to maintain the service quality of ongoing high priority communications during high load situations, while assuring that incoming prioritized connection requests receive the desired access treatment. To the best of our knowledge, this is the first article presenting a comprehensive summary of main use cases and technology opportunities for accessibility differentiation and control in 5G networks.

## II. USE CASES AND REQUIREMENTS OF 5G ACCESS CONTROL

In this section, we describe different use cases and requirements that impact the design of the accessibility differentiation framework in 5G SA networks.

### A. Prioritization of Critical Services

Due to increased spectrum needs for emerging use cases and seeking for cost-efficient mission critical (MC) solutions, more and more public safety agencies started considering use of prioritized access to mobile network operators' flexible spectrum assets, which brings the need of a proper access and admission control framework in 5G networks to support the prioritization of critical services [4].

During major emergency incidents, there can be a high-capacity demand by MC traffic to support first responders' rescue operation on site. At the same time, data traffic generated by commercial users can increase significantly. In a shared 5G network, it is crucial to ensure the flow of critical information no matter how busy the network is. This brings stringent requirements for the 5G system to be able to early identify and prioritize access requests from MC users and high priority services[5]. There might be situations where all commercial users are blocked and removed from the network, but the limited system capacity is still not enough to serve all requested MC communications. The access control mechanism plays then an important role to wisely select which of the prioritized MC users and services shall be able to access the system in agreement with the operator's policy. For this, priority differentiation between different MC users and between different MC services are needed [5].

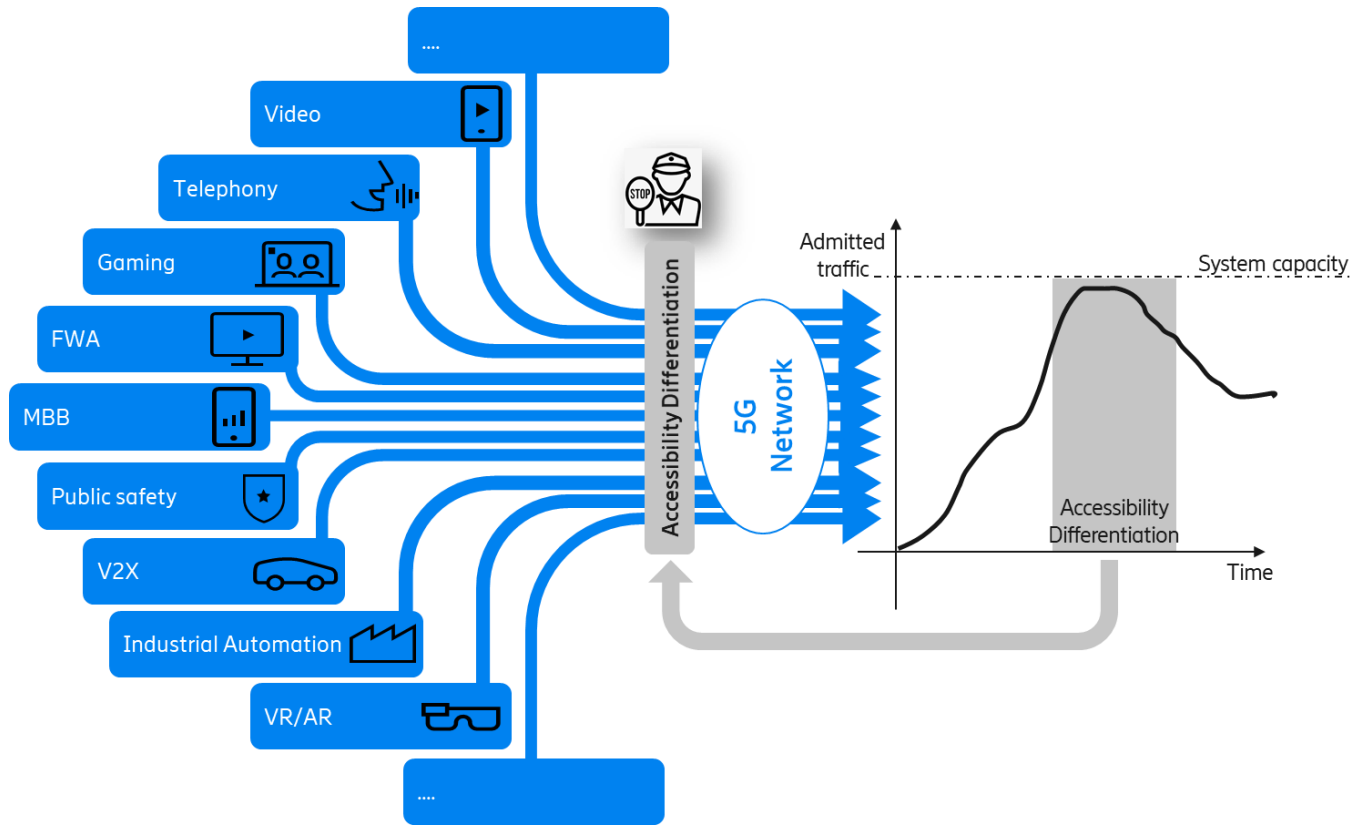

Figure 1 Example of use cases supported by 5G networks, where accessibility differentiation and control mechanisms triggered in response to high traffic load condition in the system.

*B. Non-Public Networks for Industry Connectivity*

For some industry IoT use cases such as factory automation, the main interest is the use of 5G connectivity to support customized services for a limited group of users/devices. For this purpose, the concept of non-public networks (NPNs), also known as private networks, has been introduced in 3GPP Rel-16 [3], [6]. A 5G NPN is a network that can be used by a limited group of users associated to an organization such as an enterprise, and it typically provides private 5G connectivity in a confined geographical area, such as a campus or a factory.

3GPP supports two different NPN deployment models [4]: *Standalone NPN*, which is deployed on the organization's defined premises as an independent network; and *Public network integrated NPN*, where the NPN and the public network either share part of network infrastructures for the defined premises, or the NPN is hosted entirely by the public network operators, e.g., by means of a dedicated network slice. In any of these deployment models, access control is important to ensure the NPN is only accessible for its authorized users [1].

*C. Massive Internet of Things*

Massive IoT are being used in different industry use cases such as smart traffic management, smart logistics and smart meters. In case of a server failure or a simultaneous activation of huge number of devices, all these devices will try to send their connection requests simultaneously, causing bursty traffic, which poses great challenges on access control to avoid a severe network congestion in such situations.

Massive IoT type of services typically requires only infrequent small data traffic. The signaling overhead for setting up connections before each transmission can be larger than the size of the actual data payload. To optimize the connection establishment and data transmission procedure for massive IoT with small data payloads, access differentiation is critical for the network to identify these massive IoT services at the earliest stage. This is being addressed in NR Rel-17 [7].

*D. Network Slicing Support*

The concept of *Network Slicing* has been introduced in 5G to enable operators to create multiple independent logical networks for supporting different services over a common physical infrastructure [8]. Each slice is a collection of core network (CN) and radio access network (RAN) functions and resources, customized to address the need of specific applications, users, and services. To fulfill the variety of performance demands brought by 5G-enabled services on a common 5G network, operators leverage the flexible configuration of shared and isolated pools of system resources (e.g., radio, transport, and processing resources) accessible by each slice. This poses new challenges on functionalities in charge of efficiently assign the limited system resources and differentiate between services belonging to different slices (inter-slice differentiation) and services mapped to the same slices but with different quality of service (QoS) requirements (intra-slice differentiation). The scheduler policy surely plays a key role in efficiently assigning the radio interface resources to meet the performance requirements of the traffic in the slice. However, it is the access control mechanism that can selectively admit or reject new connections, based on the possibility to serve their demands with the available per-slice resources, and provide the desired accessibility differentiation in case of resource contention between and within the slices.

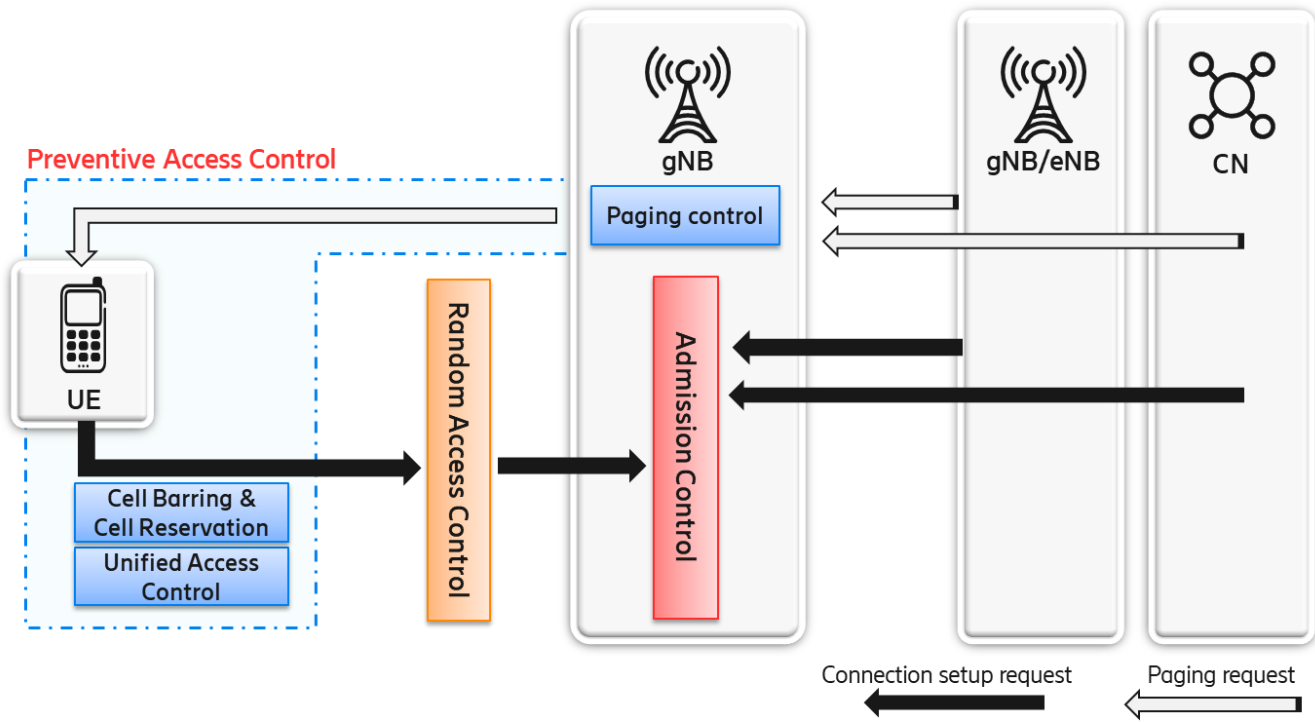

Figure 2 Overview of the 5G accessibility differentiation and control framework. Access requests coming to a given gNB either from a user equipment (UE), the CN, or other RAN nodes (gNBs or eNBs), can be selectively blocked by the activation of different mechanisms in RAN, each leveraging the available information about the request type and prioritization indicators. Here, gNBs and eNBs refer to 5G NR BSs and 4G LTE BSs, respectively.

## III. 5G Accessibility Differentiation and Control Framework

In this section, we present an overview of the 5G accessibility differentiation and control framework. As illustrated in Figure 2, the framework is divided into three main functionalities, i.e., the Preventive Access Control, the Random Access Control, and the Admission Control [9]-[11]. In case of resource limitation, each of these functionalities selectively limits the incoming traffic to the system, leveraging different connection information and tools from the 3GPP standard.

The accessibility differentiation and control actions can apply to all 5G UEs radio resource control (RRC) states [11]: RRC_IDLE, RRC_INACTIVE, and RRC_CONNECTED. A UE in RRC_IDLE has no connection with the network and moves to RRC_CONNECTED by performing initial attach or connection establishment procedures. If the UE has no activity for a given time, it can be moved to RRC_INACTIVE state, where its connection with the RAN is removed, but the one between the RAN and the CN is kept, as well as the UE Context, for faster and efficient re-transition back to RRC_CONNECTED.

The following three sections are dedicated to the description of each of these functionalities.

## IV. Preventive Access Control

Before sending any access request to a base station (BS), a UE must first check the access barring related parameters that are periodically sent by the BS in broadcasted system information, as shown in Figure 3. Based on this information, the UE determines whether its access request is allowed or not.

In this article, we use "Preventive Access Control" to refer to those mechanisms that can restrict a UE from accessing the network at the earliest stage, i.e., before the UE sends an initial access request to the network. The Preventive Access Control mechanisms can be divided into two groups [9]: the first group of mechanisms use basic system information for cell barring and reservation in order to control UE's cell selection and reselection procedures; the second group of mechanisms enables blocking new connections towards overloaded cells directly at their origin in the RAN, that is, either at the UE side for mobile originated (MO) requests, or at the BS side for paging requests. Comparing to the mechanisms in the first group, the second group enables more finer accessibility differentiation based on the user subscription and service types.

### A. Cell Barring and Reservation

The parameters related to cell barring and cell reservation are indicated in the basic system information carried on the master information block (MIB) message and the system information block 1 (SIB1) message [12]. Cell barring and reservation instructions apply only to UEs looking for a cell to camp on, they do not affect connection requests coming from other network nodes, as for dual connectivity and handover.

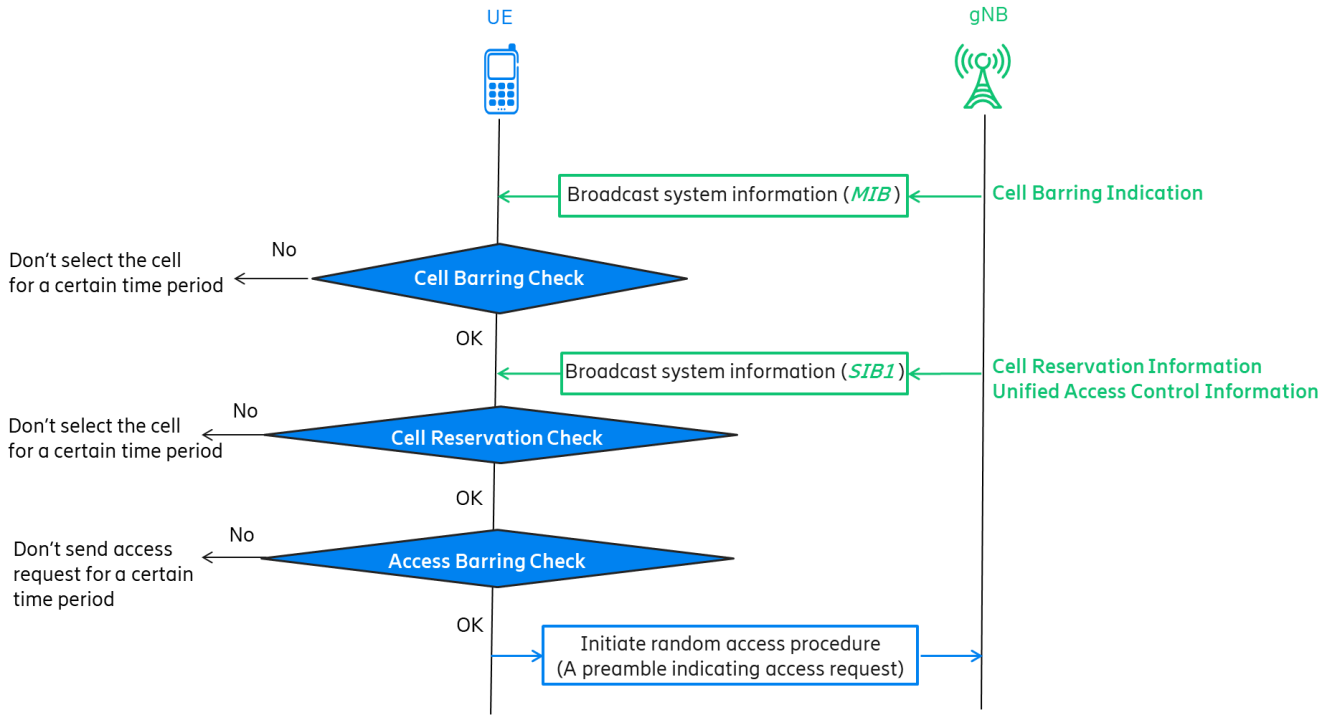

Figure 3 Illustration of the flow of 5G cell barring, cell reservation, and Unified Access Control

By means of the *cellBarred* indicator in the MIB message, it is possible to prevent any UE from selecting that cell for up to 300 seconds, not even for emergency calls. A cell can also be reserved only for specific UEs by using different indicators in the SIB1 message. The *cellReservedForOperatorUse* indicator allows only operator's users (UEs with Access Identities 11 or 15, as will be discussed in the next section) to access a given cell. This can be configured, for example, when an operator needs to do cell maintenance for a certain time period. As another example, using the indicator *cellReservedForOtherUse* together with the NPN Identity information in SIB, allows to reserve a cell for only authorized NPN UEs. 3GPP has also introduced the parameter *cellReservedForFutureUse* in the SIB1 message, as an opportunity to restrict a cell accessibility based on future 5G needs.

*B. Unified Access Control (UAC)*

UAC is the access barring mechanism introduced in NR for selectively blocking new access requests originated at the UE side [1], [9]. The advantage of UAC is that it can effectively reduce incoming traffic with no signaling and network processing consumption, as connection requests are barred before any message is sent from the UE to the BS.

In contrast to 4G long term evolution (LTE), where different access barring mechanisms were standardized to meet different needs (as described in [13]), the UAC mechanism introduced in 5G is a single framework that provides more flexibility and differentiation granularity. This is enabled by the concepts of *Access Category (AC)* and *Access Identity (AI)* described in the following. Each access attempt is associated with one AC and one or more AIs. Table 1 summarizes the AC and AI defined in 3GPP [1]. The access limitation imposed by the UAC considers also the operator serving the UEs and if the UEs are in their home public land mobile network (PLMN).

**AC:** The AC is typically configured based on the reason for the access request, i.e., the service type. NR supports 64 ACs, 32 of them are standardized, and 32 are operator-defined. In NR Rel-16, rules have been defined for mapping ACs 0 to 10 to different types of services. Standardized AC 11 to 31 are instead reserved for future use. AC 0 is mapped to *MT-access* (i.e., paging requested by the BS) and it cannot be barred. The accessibility differentiation for MT-Access is handled by network paging control instead.

An operator-defined AC can be configured with different criteria, depending on the operator needs for service differentiation. An example of such criteria is the network slice identifier, which allows to coordinate the UAC barring actions with the network slicing configuration and per-slice congestion level. Compared to standardized AC, the core network needs to inform a UE about the additional operator defined ACs supported in the network. This signaling is assumed to be done within an earlier connection to the network.

**AI:** The AIs are associated to the UE profile; that is, the information stored in universal subscriber identity module (USIM) during provisioning data. NR Rel-16 supports 16 standardized AIs. AI 0 corresponds to LTE Access Classes 0 to 9, and it is assigned to normal UEs. AIs 11 to 15 are mapped directly from LTE Access Class 11 to 15, and they are assigned for high priority UEs, such as UEs for network maintenance (PLMN-related users), in public utilities or for providing security or emergency services. AIs 12, 13, 14 are only valid for use in the home country; and AIs 11 and 15 are only valid for use in the Home PLMN (or equivalent one), indicating the operator holding the subscriber profile.

Table 1 Illustration of 5G Access Identities and Access Categories.

| Access Category (AC) | Service Type |
|---|---|
| 0 | Response to paging (MT-Access) |
| 1 | Delay-tolerant service |
| 2 | Emergency |
| 3 | Mobile originated (MO) signaling on Non-Access Stratum (NAS) level |
| 4 | MO Voice call |
| 5 | MO Video call |
| 6 | Short message service (SMS) |
| 7 | MO-data |
| 8 | MO-signaling on RRC level |
| 9 | MO IMS (IP Multimedia Subsystem) registration signalling |
| 10 | MO exception data |
| 11-31 | Reserved standardized access categories |
| 32-63 | Operator-defined |

| Access Identity (AI) | UE configuration |
|---|---|
| 0 | Regular UE |
| 1 | UE configured for Multimedia Priority Services (MPS) |
| 2 | UE configured for Mission Critical Services (MCS) |
| 3 | UE for which Disaster Condition applies |
| 4-10 | Reserved |
| 11 | UE configured for PLMN Use |
| 12 | UE configured for Security Services |
| 13 | UE configured for Public Utilities |
| 14 | UE configured for Emergency Services |
| 15 | UE configured as PLMN Staff |

To support early network accessibility prioritization for MC users in emergency situations, AI 1 and AI 2 have been introduced in 5G for UEs configured with multimedia priority services (MPS) and mission critical services (MCS), respectively. The applicability of AI 1 or 2 is indicated by the USIM configuration, and it applies only if the UE is in the country of its Home PLMN (or equivalent one). Otherwise, the applicability of AI 1 or 2 is provided to the UE by the core network in an earlier attach [14].

The UAC is applicable to UEs in all RRC states when initiating a new access attempt. When performing the UAC check, a UE maps its access attempt to an AC and one or more AIs based on the defined mapping rules. It then checks if its request is barred for the given cell by evaluating the barring information sent in the SIB1 message. For each AC, the barring information includes a *barring factor*, a *barring time*, and a *barring indicator* for each of the AIs. The barring indicator for each AI informs whether the access attempt is allowed for this AI. The barring factor indicates the probability that the given access request can be allowed. The barring time defines the minimum time interval before a new access attempt can be performed after it was barred.

Depending on the congestion level, the network can dynamically configure the barring information for each request attempt type, taking also into account the desired accessibility prioritization by the operator.

*C. Paging Control*

When new data is to be sent to a UE in RRC_IDLE or RRC_INACTIVE state, paging procedures shown in Figure 4 are triggered to get the UE to RRC_CONNECTED state.

The CN sends the *Paging Message* to the gNBs with cells belonging to the relevant Tracking Areas for the UE to be found. If new data needs to be delivered to a UE that is in RRC_INACTIVE state, the CN forwards the packets to the gNB where the UE context is still stored. This gNB then pages the UE in the cells belonging to its RAN notification area (RNA). If the RNA includes cells of neighbor gNB(s), a *RAN Paging Message* to neighbor gNB(s) is sent. Once a UE detects that it has been paged, it triggers the random access procedure and sends the connection setup request with the establishment cause set to "*MT-access*".

As mentioned earlier, *MT-access* requests are mapped to AC 0 and cannot be barred by the UAC. A UE is not aware of the paging reason, since each MT-access request can embody different connection types (e.g., simple signaling, increased number of push notifications from over-the-top applications, or high priority emergency callback). In case of overload, the network shall control the amount of incoming MT-access requests by means of Paging Control instead, which can limit the paging requests sent to the UEs and thus the corresponding MT-access requests.

Paging Control can differentiate the paging requests based on their *Paging Priority* information, which is set in the CN and provided in the *(RAN) Paging Message.* Paging Priority can be used to indicate to the Paging Control function if the packets waiting to be delivered to the UE are for prioritized services, and thus should be prioritized during high load.

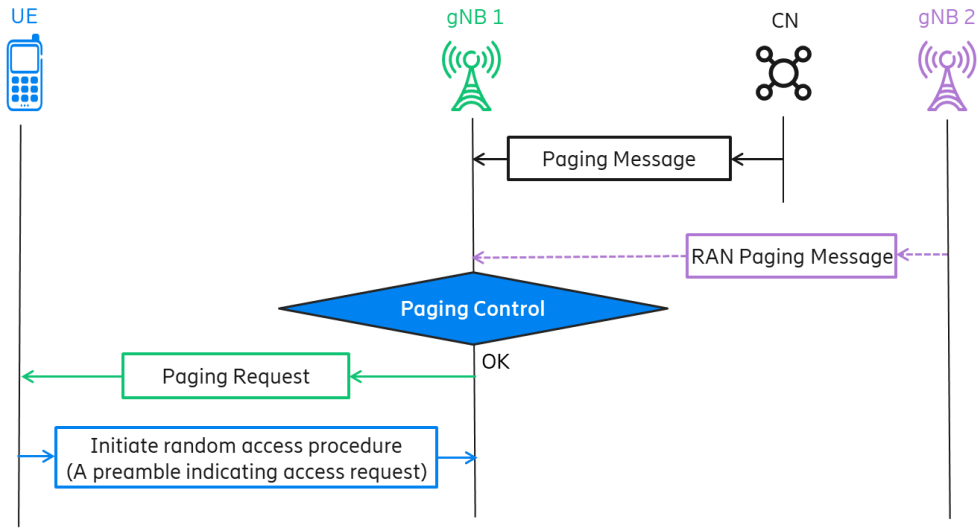

Figure 4 Illustration of the flow of 5G Paging Control

## V. RANDOM ACCESS CONTROL

If a UE is not barred and can send a connection request to the network, then, it performs the contention-based random access (CBRA). 5G NR supports two types of CBRA: *4-step CBRA*, introduced in NR Rel-15 and similar to the one adopted in LTE, and *2-step CBRA* introduced in NR Rel-16 [11]. Therefore, a 5G network can select the type of CBRA to best support the target use cases and deployment scenarios.

Figure 5 (a) shows the 4-step CBRA. A UE initiates an access attempt by transmitting a preamble (MSG1) on a physical random access channel (PRACH). After detecting the preamble, the gNB responds with a random access response (RAR) message (MSG2) on a physical downlink shared channel (PDSCH). Upon successful reception of the RAR, the UE transmits on a physical uplink shared channel (PUSCH) the MSG3 for UE identification and RRC connection request. Finally, the gNB sends the MSG4 for contention resolution.

The 2-step CBRA is shown in Figure 5 (b). In the first step, a UE transmits a MSGA, which is a combination of a preamble on PRACH and a payload on PUSCH. In the second step, the gNB responds by transmitting a MSGB, which contains the response for contention resolution. The reduced number of steps for handshaking between a UE and a gNB can provide benefits of reduced access latency and signaling overhead. For cells operating on unlicensed spectrum, the UE and gNB are typically required to perform listen-before-talk (LBT) to determine whether the spectrum channel is free or busy and transmit only if the channel is sensed free. The 2-step CBRA can reduce the number of LBT attempts, thus further reducing the access latency and increasing the spectral efficiency in unlicensed spectrum operations. However, due to possible large uplink timing errors for the MSGA PUSCH transmission, in practice, the 2-step CBRA is limited to small cell scenarios.

If a UE detects a MSG2 but fails to find the corresponding RAR matched to its preamble transmission, the UE can send a new access attempt according to the information signaled by the network [15]. In particular, the *back-off indicator* field contained in MSG2 indicates for how long a UE shall wait before retrying. The UE sets a random back-off time according to a uniform distribution between 0 and the value indicated by the *back-off indicator*. To increase the probability of success for each new RA attempt, if the UE selects the same beam (transmit antenna configuration) as for the previous attempt, it transmits the preamble with an increased power, according to the *preamble power ramping step size* indicated in SIB1 message.

To increase the likelihood of a faster successful completion of CBRA during beam failure and handover, NR Rel-15 introduced the *prioritized random access* feature [15]. This allows the UE to send a new access attempt with a shorter back-off time and a larger preamble power ramping step size, according to the configurations signaled in SIB1 message. In NR Rel-16, this feature has been further extended to prioritize network accessibility for users configured for critical services, i.e., MPS or MCS, which is not supported in 4G LTE.

## VI. ADMISSION CONTROL

If the limited system resources are reaching their maximum utilization, admission control is another functionality able to limit the incoming traffic in the system. This can be achieved by different actions, such as rejecting the incoming connections requests, queuing them to handle temporary peak of load, and/or attempt to remove already established connections to release the needed resources for incoming ones. To further regulate the amount of connections reaching a highly loaded cell, admission control can also include the optional *waitTime* information in the RRC Reject and RRC Release messages sent to those UEs not allowed to get connected [12]. Those UEs then bar all ACs, except 0 and 2, for the indicated time interval. Examples of connection procedures that can trigger admission control actions are the initial connection setup, resume, handover, and QoS flow setup request. Admission control, in fact, can act upon both connecting new users and establishing

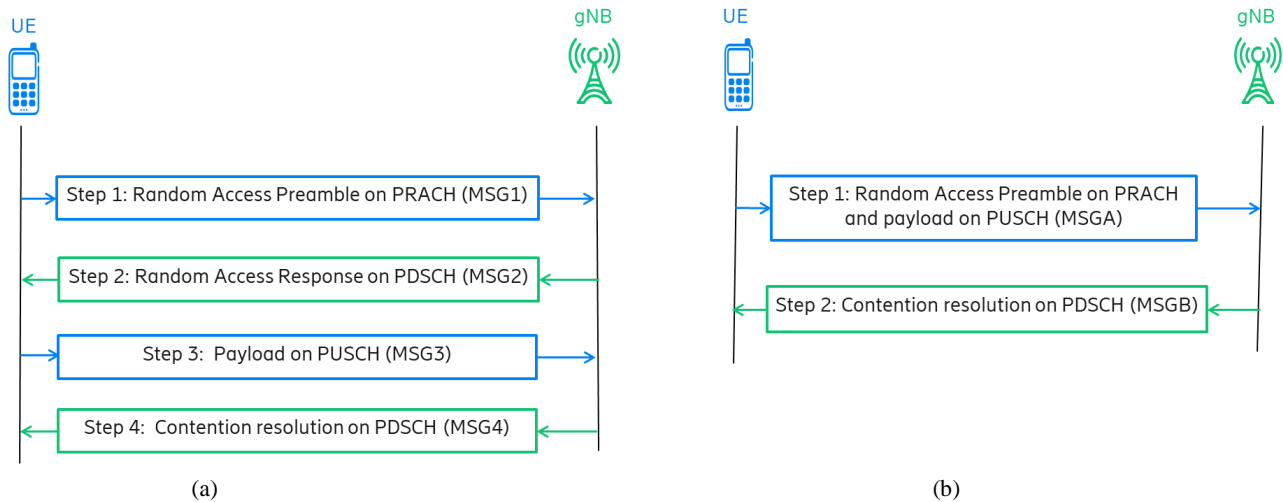

Figure 5 Illustration of 5G contention based random access (CBRA) procedures: (a) 4-step CBRA; (b) 2-step CBRA

each of the multiple services, mapped to QoS flows, that a connected user might run.

As a variety of users and services coexist in 5G networks, it is essential that admission control is capable to identify and prioritize the connection requests according to operators' choices. In what follows, we describe different standardized subscriber and service indicators that can be used by admission control to provide accessibility differentiation in 5G networks.

**Establishment Cause.** When a UE performs random access procedure for initial access, as illustrated in Figure 5, it includes the *EstablishmentCause* information in the MSG3 or MSGA PUSCH transmission, which indicates the reason for its access request [12]. To support early identification of prioritized services, like IMS-based video calls, MPS subscribers, and MCS subscribers, a set of new establishment causes, including *mo-VideoCall*, *mps-PriorityAccess* and *mcs-PriorityAccess,* has been added in NR to those already supported in LTE, like *emergency* and *highPriorityAccess*. By receiving the establishment cause at the CBRA procedure, the network can identify the priority of the associated request and decide whether to accept it, and send an RRC Setup message to the UE, or to reject it, and send an RRC Reject message instead.

**Allocation and Retention Priority (ARP).** It is a QoS parameter that defines three important prioritization aspects of a QoS flow [3]: the *ARP priority level* indicating the relative importance of the service mapped to the QoS flow (the range is 1 to 15 with 1 as the highest level of priority); the *pre-emption capability* indicating if the QoS flow can trigger the release of another one, and the *pre-emption vulnerability,* indicating if the QoS flow can be pre-empted by another one. The ARP information can be used by admission control during QoS flow establishment, modification, and handover procedures to decide, in case of resource shortage, if an incoming QoS flow setup must be rejected or can trigger the pre-emption of other ongoing connections with lower priority, to free up resources. It is worth mentioning that although ARP indicators drive QoS flows pre-emption, accessibility differentiation can also be achieved via device/subscriber pre-emption. In this case, the whole UE context is released to allow higher priority devices/subscribers to get connected.

**5G QoS Identifier (5QI).** 5QI parameter is a scalar associated with each QoS flow and used by the network as indication of the traffic forwarding treatment [3]. Standardized or pre-configured QoS characteristics are indicated through the 5QI value. Among these characteristics, the *Resource Type* categorizes the QoS flows as guaranteed bit rate (GBR), Delay-Critical GBR, and non-GBR flows. The distinction between GBR and non-GBR traffic already exists in LTE, while the Delay-Critical GBR Resource Type has been introduced in NR to support more delay-sensitive applications, such as intelligent transport systems, and automation [3].

The 5QI can thus be another valuable input for QoS flow admission decisions, as it provides indications on the type of service and on how stringent its requirements are. The GBR types of services need network resources to be consistently allocated during their lifetime to guarantee their stricter performance needs than services using non-GBR QoS flows, which are more tolerant to experience congestion-related packet drops and delays.

**PLMN**. It indicates which operator has been selected to serve the UE. A PLMN id consists of the mobile country code (MCC) and mobile network code (MNC), indicating the country and the operator of the subscription, respectively.

This info can be used both in the UAC and in admission control as subscriber differentiator. For example, to control roaming, or in case of network sharing agreement between different operators, such as the multi-operator core network (MOCN) and the multi-operator RAN (MORAN) approaches. In the latter example, the PLMN id can be used to regulate the network access to meet the sharing configuration.

**Single Network Slice Selection Assistance Information (S-NSSAI).** Each network slice is identified by an S-NSSAI, which is a combination of a mandatory slice/service type (SST) field, identifying the slice behavior in terms of features and services, and an optional slice differentiator (SD) field, which differentiates among slices with the same SST field [3].

The S-NSSAI indicator is associated with the protocol data unit (PDU) session hosting the data traffic mapped to the specific slice, and a given UE can access up to 8 network slices simultaneously via 8 distinct PDU sessions. During the initial

8context setup and the request to setup new QoS flows mapped to given PDU Sessions, the RAN is informed of the slice(s) for which resources are being requested. Admission control behavior then depends on the resource availability for the specific slice(s), and the slice-specific radio resource management policies; for example, to support slice isolation and avoid that overloading one slice impacts the delivered performances in other slices [9].

## VII. CONCLUSIONS AND FUTURE WORK

Accessibility differentiation is vital for serving multiple use cases in a shared 5G system and for securing prioritized and critical connections in various network load situations. In this article, we provided an overview, based on the 3GPP 5G standardization, of the currently supported tools for selectively limit the traffic load in the network.

Compared to 4G, 5G has introduced a set of new features and service/subscriber identifiers to support early and flexible accessibility differentiation, including

- Flexible configuration of network slices and slicing isolations to address the requirements of specific services and applications.
- The UAC framework for access barring.
- Prioritized random access to enable faster connection setup for users configured with critical services, i.e., MPS or MCS.
- Newly added establishment causes for early identification of prioritized services.
- Newly added delay-critical GBR resource type to better support delay-sensitive applications.

Further evolution of 5G accessibility differentiation and control can be envisioned. One possible direction is to enhance the support of early identification of critical services/users in random access based on e.g., differentiated PRACH configurations for different user/service priority groups. Another interesting direction is to enhance the initial access mechanisms to support a finer differentiation granularity for users/services belonging to the same service type, e.g., MCS. Also, to efficiently use the accessibility differentiation tools described in this article, it is important to understand the design rationales and coordinate their interplay. In this context, it is interesting to explore the role of machine learning in enabling automated operations that adapt to different requirements and network situations.

BIOGRAPHIES

**Jingya Li** (jingya.li@ericsson.com) is a Master Researcher at Ericsson Research, Gothenburg, Sweden. She currently leads 5G/6G research and standardization in the areas of public safety mission critical communications. She has led different back-office feature teams for 5G New Radio (NR) standardization. Her contribution to standardization has been on 5G NR for public safety, initial access, remote interference management and cross link interference handling, vehicle-to-anything communications, and latency reduction. She is the recipient of the IEEE 2015 ICC Best Paper Award and IEEE 2017 Sweden VT-COM-IT Joint Chapter Best Student Journal Paper Award. She holds a Ph.D. degree (2015) in Electrical Engineering from Chalmers University of Technology, Gothenburg, Sweden.

**Demia Della Penda** (demia.della.penda@ericsson.com) received the M.Sc. degrees in telecommunications engineering from the University of Siena, Italy, in 2012, and the Ph.D. degree from KTH Royal Institute of Technology, Sweden, in 2018. Since 2018, she has been with Ericsson AB, Stockholm, Sweden, working on the design and systemization of 5G radio access networks. Her efforts are mainly focused on features and optimized solution for QoS and handling of high load scenarios in RAN.

**Henrik Sahlin** (henrik.sahlin@ericsson.com) is research manager within the section of microwave systems at Ericsson Research, Gothenburg, Sweden. Previously he was project manager with focus on automotive, transport and public safety related research. He has been actively participated in 3GPP (Third Generation Partnership Project) standardization for LTE (Long Term Evolution) and NR (New Radio, 5G) with a focus on initial access and reduced latency with physical layer design. Receiver algorithms in base stations and mobile devices has been a central research area for Henrik within several wireless standards such as NR, LTE, WCDMA and GSM. Henrik holds a Ph.D. within the area of signal processing in electrical engineering from Chalmers University of Technology in Gothenburg, Sweden.

9**Paul Schliwa-Bertling** (paul.schliwa-bertling@ericsson.com) serves as an expert in mobile networks architecture and signaling at Ericsson Research in Linköping, Sweden. He has worked extensively with the development of RAN product and system-level concepts as well as 3GPP standardization across multiple generations of RAN and CN. His current work focuses on the evolution of network architecture and the related signaling aspects contributing to 3GPP standardization. He holds an M.Sc. in electrical engineering from the University of Duisburg-Essen in Germany.

**Mats Folke** (mats.folke@ericsson.com) is a Master Researcher at Ericsson, Sweden. With more than 10 years of experience of being a standardization delegate in 3GPP RAN WG2, his efforts span multiple access technologies and features. He has focused on device-to-device features in LTE and user plane protocols in NR. Mats received an M.Sc degree in computer science and engineering (2003) and a Licentiate degree in computer communication (2006) from Luleå University of Technology, Sweden.

**Magnus Stattin** (magnus.stattin@ericsson.com) is a Principal Researcher at Ericsson. He graduated and received his Ph.D. degree in radio communication systems from the Royal Institute of Technology, Stockholm, Sweden in 2005. He joined Ericsson Research in Stockholm, Sweden, in June 2005. At Ericsson Research he has been working with research in the areas of radio interface and radio network protocols and architecture of various wireless technologies. He is active in concept development and 3GPP standardization of LTE, LTE-Advanced, NB-IoT, NR and future wireless technologies.